\documentclass[onecolumn,amsmath,amssymb,pra,superscriptaddress,a4paper,floatfix,showkeys]{revtex4}
\usepackage{graphicx}
\usepackage{mathptmx}      % use Times fonts if available on your TeX system

\begin{document}

\title{Hidden assumptions in the derivation of the Theorem of Bell\footnote{Invited paper presented at FQMT11}}

\author{Karl Hess}
\email{k-hess@illinois.edu}           %  \\
\affiliation{%
Beckman Institute, Department of Electrical Engineering and Department
of Physics, University of Illinois, Urbana, Il 61801, USA
}%
\author{Hans De Raedt}
\email{h.a.de.raedt@rug.nl}
\affiliation{%
Department of Applied Physics,
Zernike Institute for Advanced Materials,
University of Groningen, Nijenborgh 4, NL-9747 AG Groningen, The Netherlands
}%
\author{Kristel Michielsen}
\email{k.michielsen@fz-juelich.de}           %  \\
\affiliation{%
Institute for Advanced Simulation, J\"ulich Supercomputing Centre,
Research Centre J\"ulich, D-52425 J\"ulich, Germany
}%

\begin{abstract}
John Bell's inequalities have already been considered by Boole in 1862.
Boole established a one-to-one correspondence between experimental outcomes
and mathematical abstractions of his probability theory.
His abstractions are  two-valued functions that permit the logical operations AND, OR and NOT
and are the elements of an algebra.
Violation of the inequalities indicated to Boole an inconsistency of definition of the abstractions
and/or the necessity to revise the algebra.
It is demonstrated in this paper, that a violation of Bell's inequality by Einstein-Podolsky-Rosen type
of experiments can be explained by Boole's ideas.
Violations of Bell's inequality also call for a revision of the mathematical abstractions and corresponding algebra.
It will be shown that this particular view of Bell's inequalities points toward an incompleteness of quantum mechanics,
rather than to any superluminal propagation or influences at a distance.
\end{abstract}

\date{\today}

\maketitle

\section{Introduction}

We discuss Bell's inequalities \cite{bell} and violations of them in terms of the work of Boole \cite{boole}. Boole had derived
inequalities similar to those of Bell more than 100 years before Bell, and had traced their violation to an incorrect
definition of the mathematical abstractions that represent experimental outcomes. We have shown previously that violations of
Bell's inequalities can be interpreted in a similar fashion \cite{hmd,hmd1,hmd2}. We show here further, that Bell's own interpretation of
violations of his inequalities is based on several unwarranted assumptions. For example, he assumed that his mathematical abstractions
automatically follow the algebra of real numbers and he did not permit an explicit time dependence of his functions.
Permitting such explicit time dependence leads us to a commutative algebra involving stochastic processes.
These processes invalidate what we call Bell's impossibility proof: Bell's notion that ``quantum mechanics can not be embedded
in a locally causal theory" \cite{cuisine}. Our findings favor the suggestions of ``incompleteness" as described in the
Einstein-Podolsky-Rosen (EPR) \cite{EPR} paper. These findings confirm and extend numerous mathematical and physical treatises
see e.g. \cite{vaxjo,KHRE08,KHRE09,NIEU11}.

\section{Possible experimental outcomes and Boole's algebra}

Boole \cite{boole} discussed a mathematical-logical way to deal with statistics and probabilities. He dissected experiments into
events that could only assume two values obeying a calculus that resembled the algebra of real numbers, but with the operations
of multiplication, addition and subtraction replaced by the respective logical operations of conjunction, disjunction and
negation. The experimental results were replaced by these mathematical abstractions, as soon as a valid one-to-one
correspondence between the experiments and the abstractions was established. Consider, for example patients in the two cities A
and B, who have certain different histories denoted by the bold faced letters $\bf{a, b, c...}$. These patients are examined
with regard to a certain disease and the results of the examination are filed in the following way. If a patient from city A
with history $\bf a$ tests positive a note is made that $A_{\bf a} = +1$ and if the test is negative we have $A_{\bf a} = -1$.
If the patient is from city B, then we have $B_{\bf a} = \pm 1$ etc.. The subscripts $\bf a$, $\bf b$ etc. indicate, for
example, different levels of fever that the patients have encountered.

Boole realized, and this is crucial, that the mathematical abstractions that are chosen in correspondence to the experiments are
not necessarily elements of an algebra. It is possible, that the disease that is investigated in the above example depends on
factors other than the fever of the patients and their place of residence. It could also depend on the date of their birth.
Boole, therefore, tried to establish criteria that could be used in order to determine whether the chosen correspondence of
experiments and mathematical abstractions was a valid one, and whether the abstractions therefore did follow the rules of logic.
His idea was to employ the algebra of real numbers to the mathematical abstractions, to deduce in this way non-trivial
inequalities and to see whether these inequalities were consistent with all experimental results; in our example with the data
from the patients. One type of inequality that he considered was:
\begin{equation}
A_{\bf a}A_{\bf b} + A_{\bf a}A_{\bf c} + A_{\bf b}A_{\bf c} \geq -1 .\label{11may23n1}
\end{equation}
The fact that this sum of products is larger than or equal to $-1$ can easily be checked by inserting all possible values of
$\pm 1$. The inequality is non-trivial, because knowing nothing about the experiments that are described, and realizing that all
the measured outcomes may derive from different persons, a possible result is $-3$.  This can immediately be seen by adding 6
different birth dates as a superscript. Then we clearly can have:
\begin{equation}
A_{\bf a}^1A_{\bf b}^2 + A_{\bf a}^3A_{\bf c}^4 + A_{\bf b}^5A_{\bf c}^6 \geq -3. \label{11may23n2}
\end{equation}

Thus, dissecting any experiment into binary functions, does not necessarily result in mathematical abstractions that follow the
algebra of real numbers. For a given sequence of experiments, and definition of corresponding mathematical abstractions, one
needs to make sure that these indeed follow an algebra in order to certify that the experiments were ``dissected" sensibly.
Boole's idea was therefore to take statistical averages of expressions as given in Eq.(\ref{11may23n1}). If a violation of the
corresponding inequality for the expectation values was obtained, then that was an indication that the experiment could not be
fully understood in terms of the assumed mathematical abstractions. Different abstractions must then be chosen to correspond to
the experiments, e.g. those shown in Eq.(\ref{11may23n2}) including the birth dates.

\section{Vorob'ev's cyclicities}

Boole's approach to probability theory can be seen as a special case of Kolmogorov's framework that deals with $\sigma$-algebras
of countable sets of events. A great review of Kolmogorov's probability theory with special emphasis on physics experiments has
been given in \cite{kolmo}.

Progress related to Boole's consistency tests with inequalities was made by Vorob'ev \cite{vorob}. He showed, in a very general
way, that non-trivial inequalities and conditions of the Boole type can be found by constructing topological-combinatorial
``cyclicities" of functions on $\sigma$-algebras. For the purpose of our paper, it is sufficient to understand these general
cyclicities just by the above example: the $A_{\bf i}A_{\bf j}$ with $ {\bf i, j} = {\bf a, b.c}$ are functions on a probability
space (random variables) and form a closed loop, meaning that the choices in the first two products of the inequality determine
the third. An infinite number of such inequalities can, therefore, be composed by arranging algebraic expressions of functions
that determine the value of other algebraic expressions of the same inequality. Any violation of such inequalities by the
measured averages means that the mathematical abstractions describing the experiments are not functions on a $\sigma$-algebra.

Indeed, mathematical-logical abstractions constructed for experiments, by using labels that are derived from sense impressions
(such as $A_{\bf a}$), do not necessarily follow the algebra of Boolean variables, the algebra of real numbers or any other
given algebra. However, within the framework of special relativity, one can always find consistent abstractions
because of the following general reason. Each experiment, and each mathematical abstraction, an event as
defined by Kolmogorov \cite{kolmo}, corresponds to a different space-time coordinate $st_n = (x_n, y_n, z_n, t_n), n = 1, 2,
3...$, or set of such coordinates, that represent events as defined by Einstein in his special relativity for some given
inertial system. Therefore, no matter how many experiments and cyclicities we consider, the outcomes of the experiments that we
register and record can always be labeled by different space-time labels $st_1, st_2,..., st_n, ...$. We can, in general,
rewrite the example of Eq.(\ref{11may23n1}) as: \begin{equation} A_{\bf a}^{st_m}A_{\bf b}^{st_{(m+1)}} + A_{\bf
a}^{st_{(m+2)}}A_{\bf c}^{st_{(m+3)}} + A_{\bf b}^{st_{(m+4)}}A_{\bf c}^{st_{(m+5)}} \geq -3, \label{11may23n3} \end{equation}
and thus have removed the cyclicity for all inequalities with $m = 1, 7, 13, 19,...$. The functions that we use have become more
numerous, but they do follow the algebra of real numbers, and no contradiction of the Boole-, or Vorob'ev- type can be found,
because all inequalities based on topological combinatorial cyclicities can be removed that way. Note that the setting labels
$\bf a, b, c...$ and space-time labels $st_m$ are used as indices and not as independent variables. As will be discussed in more
detail, this is important because certain settings can not occur at certain space-time coordinates, and these two variables are
therefore not independent.

\section{Bell's inequality}

More than 100 years after Boole, an inequality similar to the inequality of Eq.(\ref{11may23n1}) was re-discovered by John Bell
\cite{bell1}, who analyzed quantum experiments that were constructed to investigate work of Einstein, Podolsky and Rosen (EPR)
\cite{EPR}. Bell did not refer to Boole's work, nor to the contemporary work of Vorob'ev in any of his collected papers
\cite{bell}, and probably did not know of them. He considered measurements of correlated spin-pairs at two
locations (cities) A and B. The results of these measurements involve two spin outcomes ($A, B = \pm $1), one on each side.
Quantum mechanics predicts that for certain instrument settings $\bf a, b, c....$ the experimental outcomes obey $A_{\bf i} = -
B_{\bf i} , {\bf i} =, {\bf a}, {\bf b}, {\bf c}....$, where $\bf a, b, c...$ are now three-dimensional unit vectors related to
the settings of the spin-measurement equipment. Bell also introduced a variable $\lambda$ to be discussed in detail below and
constructed from these facts an inequality equivalent to:
\begin{equation} A_{\bf a}(\lambda)B_{\bf b}(\lambda) + A_{\bf
a}(\lambda)B_{\bf c}(\lambda) + A_{\bf b}(\lambda)B_{\bf c}(\lambda) \leq +1 .\label{11may25n1}
\end{equation}
This is again a nontrivial inequality and follows from Boole's corresponding equation by replacing some of the A's by B's and multiplying by
$-1$. The equation is non-trivial, because in general the result could be as large as 3. This equation is now often called
Bell's inequality, while Eq.(\ref{11may23n1}) is sometimes referred to as the Leggett-Garg inequality. Note that Bell used the
slightly different notation of $A({\bf a}, \lambda), B({\bf b}, \lambda)$ etc. because he treated (mistakenly, as explained
below) the settings $\bf a, b, c...$ and $\lambda$ as independent variables.

Quantum mechanics provides us also with the result that the expectation value $E_{entangled}$ for the spin correlations of
pairs of spins in the singlet state is given by:
\begin{equation} E_{entangled} = -{\bf i} \cdot {\bf j} \text{ with } {\bf i,j} =
{\bf a, b, c, ....}. \label{11may25n2}
\end{equation}
This result violates Eq.(\ref{11may25n1}), because one can find $\bf a, b,
c...$ such that each of the products in Eq.(\ref{11may25n1}) equals $1/\sqrt{2}$ resulting in a left hand side of
$3/\sqrt{2} > +2 $.

\section{Bell's explicit and implicit assumptions}

The assumptions, made by Bell in his collected works \cite{bell} to derive his inequalities, are numerous and vary in the
different papers of Bell and his followers. We mention here just some of the crucial explicit and implicit (or ``hidden")
assumptions. \begin{itemize}

\item[(i)] Bell treats $\lambda$ as an element of reality in the sense defined by EPR and proposes the hypothesis that $\lambda$
effects a more complete specification of the ``state" of the correlated spin pair.

\item[(ii)] The instrument settings $\bf a, b, c...$ and $\lambda$ are assumed to be independent variables. This is often,
mistakenly, deduced from the freedom that the experimenter undoubtedly has to choose the settings.

\item[(iii)] Bell further treats $\lambda$ as a very general mathematical variable \cite{bell1}:  ``It is a matter of
indifference in the following whether $\lambda$ denotes a single variable or a set or even a set of functions, and whether the
variables are discrete or continuous."

\item[(iv)] Bell assumes that violations of his inequality are directly connected to $\lambda$ and suggests that violations
imply influences of both measurement settings on $\lambda$.

This latter suggestion led to the well known experiments by Aspect and others \cite{Asp}, who changed the measurement settings
of both sides rapidly and more or less randomly. This rapid switching excludes certain physical influences of the settings on $\lambda$
that propagate with finite speed (slower or equal to that of light in vacuum). After having received knowledge of the
experimental results of Aspect and others that violated his inequality, Bell stated that ``quantum mechanics can not be embedded
in a locally causal theory" \cite{cuisine}.

\end{itemize}

The attributes of $\lambda$ listed in (i)-(iii) are, when taken to the limit of their stated generality, in logical, physical
and mathematical conflict with each other.

Most importantly, space-time variables such as $st_m$ and instrument settings such as $\bf a, b, c...$ are not independent
variables. In the reference frame of the laboratory, any instrument setting is related to a certain space like variable because
of the location of the instrument, and to a time like variable because there can not be two different settings at the same time
and location. As soon as the settings are (indeed freely) chosen at certain $st_m$, that coordinate-set is not available for any
other settings.

Bell states in \cite{cuisine} that ``we can imagine these settings being freely chosen at the last second by two different
experimental physicists....if these last second choices are truly free or random, they are not influenced by the variables
$\lambda$." While this statement may be true for some $\lambda$, Bell's implicit assumption that, therefore, $\lambda$ and the
settings $\bf a, b, c...$ are also mathematically and physically speaking independent variables, is false in general. If
$\lambda$ represents space-time variables, then these variables and the setting pairs are not independent.

  Quantum mechanics separates the setting related variables entirely from space-time by using settings $\bf a, b, c...$ in
connection with operators such as $\sigma_{\bf a}, \sigma_{\bf b}, \sigma_{\bf c}...$, and space-time in connection with wave
functions $\psi(st_m)$. Classical probability also can not and must not use all of these variables as independent. Indeed,
Kolmogorov's stochastic processes add a separate time index to each measurement $A, B$ and do not explicitly include equipment
settings.

  Bell's functions of both the settings and the $\lambda$'s, such as $A({\bf a}, \lambda)$ are not formulated with the
appropriate mathematical and physical caution, and are either not general or not functions of independent variables, as both
assumed in Bell's proof.  Even for purely formal reasons, Bell's $\lambda$ can not be both the independent time variable of
physics and a random variable, simply because time is not a random variable. Time is also is not an operator in quantum
mechanics!

The absorption of independent space-time variables into Bell's $\lambda$ also leads to logical contradictions for yet another
reason. The proofs of Bell and followers often assume, explicitly or implicitly, that $\lambda$ occurs in sets of six (sometimes
more) to maintain a Vorob'ev cyclicity. Bell orders the functions in his initial proof into sums of the three products as shown
in Eq.~(\ref{11may25n1}), all containing the identical variable $\lambda$. This fact looks innocuous in Bell's paper, because at
the point at which he invokes the appearance of $\lambda$ in six factors, $\lambda$ is treated as a dummy integration variable
(see the equation without number after Eq.(14) in Bell's original paper). To show the problems connected with these choices of
$\lambda$, we distinguish now two cases.

First assume that the $\lambda$'s are indeed elements of reality that play a role in the formation of the data that are actually
collected. Such $\lambda$'s are in principle all different (naturally averages over large numbers may still be the same).
Consider now such $\lambda$'s formed by a combination of a
space-time variable $st_m$ and some other arbitrary element of reality $\lambda_m'$ related to spin. Then consider the
parameters $\lambda_m = (st_m, \lambda_m')$ and pick six equal values of $\lambda_m =\lambda$, for each of the six different
measurements given in Eq.(\ref{11may25n1}). One then has equated different times and different spin-related parameters, which
may be physically unreasonable. Consider as an example the boiling of an egg and link $st_m$ to various clock-times during the
process of boiling, as suggested by H. B. G. Casimir \cite{cuisine}. Further denote the egg viscosity at various locations in
the egg ( settings, such as in the yoke or in the egg-white) by $\lambda_m'$. It certainly is incorrect for this case to equate
a variety of parameters $\lambda_m = (st_m, \lambda_m')$ for different settings. The yoke stays soft much longer than the white.
Thus, for general Einstein local experiments, $\lambda$ may have to depend on both settings and times.  Equating the $\lambda$'s
in general cases, as Bell has done for his $\lambda$'s and also his ``be-ables" \cite{bell}, can therefore lead to logical
contradictions. Space-time coordinates and other general elements of reality can not be concatenated into one variable and then
regarded as the same dummy variable for triple (or quadruple) products. In fact, if we wish to prove Bell's inequality, the
possibility of ordering all data into the triple products of Eq .~(\ref{11may25n1}) is just what needs to be shown. Bell,
obviously, did not consider the full implications of Casimir's counter-example.

Second assume with Mermin \cite{mer}, that the $\lambda$'s of Eq.~(\ref{11may25n1}) are indeed all the same and the equation really is
written, as Mermin states, only to contemplate the conjecture that there exist such $\lambda$'s that ``predetermine" the
experimental outcomes. Only one of the experiments listed in Eq.~(\ref{11may25n1}) is then being performed, according to Mermin.
However, also according to Mermin, if the outcomes are predetermined by the $\lambda$'s, one can imagine obtaining results as
written down in Eq .~(\ref{11may25n1}). This statement is principally correct, however,
such results would have nothing to do with actually collected
data of actual experiments that are used for the formation of the expectation values. Nor do the results of
Eq .~(\ref{11may25n1}) have anything to do with Kolmogorov's or Boole's probability theory that deal with possible outcomes of
experiments that can be collected as ``data" and described by sensible mathematical abstractions. The hypothesis of possible
predetermination of outcomes is irrelevant to Kolmogorov's sample space. Consider the meals on the menu of a variety of
restaurants. These are certainly predetermined. Yet, when we eat in hundreds or thousands of restaurants of type ${\bf a} =$
Indian, ${\bf b} =$ Chinese, ${\bf c} =$ Austrian, etc., and eat only one meal at a time, then we do not necessarily have the
expectation of having averages from all the meals of all the menus in our stomachs, but rather averages and corresponding
correlations arising from one meal at each place. Bell's own criticism of von Neumann's work can be directly applied here: ``It
was the arbitrary assumption of a particular (and impossible) relation between the results of incompatible measurements {\it
either} of which {\it might} be made on a given occasion but only one of which can in fact be made" \cite{bvn}. Speaking in more
precise mathematical terms, one can not apply the pointwise ergodic theorem to the functions of Eq.(\ref{11may25n1}), because
the experiments that correspond to these functions cannot be performed except for one pair of settings.
In case of a violation of the inequality, no $\sigma$-algebra exists on which these functions can be defined and, therefore,
Eq.(\ref{11may25n1}) has no consequences for the expectation values obtained in the actual experiments.

The assumption that $\lambda$ may not depend on the setting parameters $\bf a, b, c...$ in a locally causal theory is also in
conflict with the mathematical generality of $\lambda$. If we wish to regard $\lambda$ as an event in the sense of Kolmogorov,
then we need to be able to identify $\lambda$ with the event that $A$ and $B$ with chosen settings assume certain values. Thus
we need to be able to identify the $\lambda$'s with events (usually denoted by $\omega$) of a probability space. The actual
event (usually denoted by $\omega_{act}$ \cite{willi}) could, for example, be $A_{\bf a} = +1, B_{\bf b} = +1$. That actual
event certainly is not independent of $\bf a, b$. This identification has nothing to do with causal theories but only with the
fact that the event is composed of two different experiments. Whether or not influences at a distance are present is not a
concern of probability theory. If one wishes to use established probability theory and, therefore, wishes to identify a specific
$\lambda$ with a an event realization $\omega_{act}$, a logical difficulty arises.

Fortunately, it is not necessary to include all these subtle points, in order to discuss the relevance and ranges of validity
of Bell's theorem.  The following two incorrect assumptions of Bell are straightforward to understand and, together with (ii)
sufficient to show how Bell's proofs and the proofs of his followers fail.

\begin{itemize}

 \item[(v)] Bell always assumed or implied that his mathematical abstractions representing the experiments follow automatically
the algebra of real numbers. This is actually what needs to be shown by the fulfillment of all possible inequalities and other
consequences of Vorob'ev's cyclicities.

\item[(vi)] Bell treated $\lambda$ explicitly as a random variable with well defined and given probability distribution
$\rho(\lambda)$ (see his Eq.(12) in \cite{bell1}). Hidden behind this innocuous assumption is the fact that, therefore, {\it
Bell did not permit an explicit space-time dependence of the statistical properties of his functions $A, B$, and has therefore
excluded general stochastic processes!} The perception that Bell and his followers had was that $\lambda$ can represent the
space-time variables. However, as we discussed above, then the settings can not be regarded as independent variables.

\end{itemize}

Work of Bell's followers often implies assumptions and restrictions of similar nature. For example, Leggett and Garg
\cite{lgarg} and also Mermin \cite{mer} always assume that their symbols follow the algebra of real numbers. They also do not
permit, and this is our crucial point, an explicit time dependence of the statistical properties of the functions $A, B$.
Furthermore, whenever they use $\lambda$'s, they do use the settings and $\lambda$'s as independent variables
by integrating (summing) independently over them. Mermin even
claims toward the end of his paper \cite{mer} ``...times...can be independently varied without altering the distribution". At
best, however, the expectation values can stay unchanged by variations of measurement times. Because of the coincidence counting
of the actual experiments, even this statement is not of general validity. Nor does quantum theory or any physical theory tell
us that time can be varied without altering probability distributions.
Quantum theory does provide us with time dependent probability distributions and with expectation values for large numbers of measurements.
Single outcomes or small sets of outcomes are not the objects of quantum theory,
that most certainly does not forbid an explicit space-time dependence of the statistical properties of the functions $A, B$.
These functions are not even part of quantum theory.

Removing the assumptions (ii), (v) and (vi) of Bell and followers does, therefore, open new horizons. One can envision the
violation of Bell-Boole type of inequalities, or more generally, Vorob'ev cyclicities as prescriptions for necessary space-time
dependencies and, therefore, as physical rules that are not contained in quantum theory. One can then construct explicitly
space-time dependent functions on $\sigma$-algebras that remove all Vorob'ev type cyclicities that lead to contradictions. These
functions describe then a more complete physical theory in the sense of EPR.

\section{Removing Bell's assumptions: generalized stochastic processes}

Instead of asking which properties of $\lambda$ might explain a violation of Eq.(\ref{11may25n1}), the more general and crucial
question to ask is: under which circumstances it is possible to remove the cyclicity in Eq.(\ref{11may25n1}) in order to
obtain the quantum result of Eq.(\ref{11may25n2}) within the Kolmogorov framework? This can indeed be done by using a slightly
generalized version of a stochastic process.

We consider here only a discrete time stochastic process. Such a process is defined by a finite or countable infinite sequence
of random variables such as $A_{\bf a}^{t_1}, A_{\bf a}^{t_2}, A_{\bf a}^{t_3}, ....$. We add the slight generalization that we
use space-time coordinates instead of time coordinates and thus have $A_{\bf a}^{st_1}, A_{\bf a}^{st_2}, A_{\bf
a}^{st_3}, ....$. This generalization is trivial as long as we deal with discrete time stochastic processes, because then the
physical meaning of the time related index makes no difference for the mathematics. For Einstein-Podolski-Rosen experiments, we
also introduce a second stochastic process $B_{\bf b}^{st_1'}, B_{\bf b}^{st_2'}, B_{\bf b}^{st_3'}, ....$ for the correlated
pair ($A_{\bf b}^{st_1'}=-B_{\bf b}^{st_1'}$ etc.),
and note that the primed and unprimed space-time coordinates are correlated in such experiments by the technique of
coincidence counting of the pair results. Combining the two stochastic processes, one obtains a vector stochastic process $
(A_{\bf a}^{st_1}, B_{\bf b}^{st_1'}), (A_{\bf a}^{st_2}, B_{\bf b}^{st_2'}), (A_{\bf a}^{st_3}, B_{\bf b}^{st_3'}), ...$.

The vector stochastic process as defined above can be further generalized by choosing the setting pairs (indicating the
measurement procedure) randomly. Thus we obtain the discrete space-time vector stochastic process:
\begin{equation}
(A_{\bf i}^{st_1}, B_{\bf j}^{st_1'}), (A_{\bf i}^{st_2}, B_{\bf j}^{st_2'}), (A_{\bf i}^{st_3}, B_{\bf j}^{st_3'})
\text{...  with }{\bf i, j} = RP,
\label{11may25n3}
\end{equation}
where RP denotes any pair $({\bf a}, {\bf b})$, $({\bf a}, {\bf c})$ or $({\bf
b}, {\bf c})$ chosen totally randomly, or at the will of any experimenter.  This is indeed a vector stochastic process. The only
difference to the standard definition is that the discrete times $t_n$ are replaced by discrete space-time labels $st_n$ with $n
= 1, 2, 3,...$. Furthermore, certain setting pairs are chosen at the discrete space-times to result in particular pairs of
functions on a probability space $(A_{\bf i}^{st_n}, B_{\bf j}^{st_n'})$ that are usually just denoted by pairs of functions
such as $ X^nY^n$, with both $X^n$ and $Y^n$ being functions on a probability space $\Omega$ with elements $\omega$
\cite{willi}.

Care must be taken with the choice and labeling of the functions. It is not possible, for example, to construct a four
dimensional vector stochastic process with equal space-time labels of the vectors that returns the quantum result. This is
because such a process also fulfills the nontrivial inequality
\begin{equation} (A_{\bf a}^{st_n} B_{\bf b}^{st_n}) + (A_{\bf
a}^{st_n} B_{\bf c}^{st_n}) + (A_{\bf b}^{st_n} B_{\bf c}^{st_n}) \leq +1
,
\label{11june7n1}
\end{equation}
that involves a cyclicity. It also contains the fundamental error to assign equal space-time labels to different equipment settings, thus
ignoring that settings and space-time labels are not independent variables.

The fact that processes as given by Eq.~(\ref{11may25n3}) remove any cyclicity, can be seen immediately by inserting the pairs
into Eq.(\ref{11may25n1}). Because all factors are in principle different. The resulting inequalities are of the trivial form:
\begin{equation}
A_{\bf a}^{st_m}B_{\bf b}^{st_m'} + A_{\bf a}^{st_{(m+1)}}B_{\bf c}^{st_{(m+1)}'} + A_{\bf b}^{st_{(m+2)}}B_{\bf
c}^{st_{(m+2)}'} \leq +3
,\label{11may26n1}
\end{equation}
for $m =1, 4, 7,...$.

It can be shown that for any given stochastic process of the type Eq.~(\ref{11may25n3}), with randomly chosen settings and
space-time coordinates, one can find a probability measure that indeed leads to the quantum result \cite{hp1}. That probability
measure must necessarily depend in some complex way on the setting pairs. This fact has a mathematical reason that is
independent of any locality considerations.The settings and space-time variables are, as stressed above, not independent. For
each setting pair, one has necessarily a different function-domain of space-time variables $st_n$. Therefore, if the setting
pairs change, the domain must also change, and so must, in general, the corresponding joint probabilities for the functions $A,
B$. These probabilities depend, therefore, on the setting pairs. Probability theory can, thus, not ``play" the so called Bell
game.

The Bell game requires two players in two measurement stations to choose values for $A, B$ with the knowledge of the instrument
settings of their respective stations only. The expectation values for these choices of $A, B$ need then to give the quantum
result. However, any probability model of the game needs to involve, as shown above, different domains of the functions for
different setting pairs. If the Bell inequalities are fulfilled, this does not matter because then one can find one
``artificial" probability space that leads to the experimental expectation values. If, on the other hand, the Bell inequalities
are not fulfilled, then no such probability space exists. The $A, B$ need to be labeled with a space-time index to avoid cyclicities,
and the player needs to know the actual domains of the functions and actual joint probabilities to
play the game. These domains and joint probabilities can not be determined from the knowledge of one setting in one station only.
Thus there exists a purely
mathematical reason for the impossibility to play the Bell-game with probability theory. To play the game by other means would
then require necessarily a detailed knowledge of all possible space-time dependencies of measurement and preparation in EPR
experiments, and thus an infinity of functions $A^{st_n}, B^{st_n'}$ with the cardinality of the continuum \cite{hp2}.  We have
not proven the existence of such functions by the above findings but we have refuted the impossibility-proofs of Bell and
followers because we removed the cyclicities by use of general space-time labels. No assumption needs to be made, from the
viewpoint of logic or mathematics, whether these space-time labels are from within or from outside the light-cone.

\section{Space time dependencies of spin measurements}

It remains to be shown that space-time dependencies of the preparation of correlated pairs as well as the measurement outcomes
(and therefore of the functions $A, B$) are physically reasonable, and can lead to the quantum result of Eq.~(\ref{11may25n2}).

Possible time dependences within the light cone are numerous. We just list here a smorgasbord of those that matter for sets of
particles with spin.  The earth rotates around itself and this rotation introduces a time dependence on how a gyroscope would
be seen from a resting observer. Naturally that time dependence would be of the order of a day. Radio waves can cause spin
resonances and are omnipresent as are magnetic fields. Because the earth magnetic field is small, corresponding time
dependencies would be slow also. The interaction of nuclear spins and electron spins happen in a wide range and equilibration of
electron spin in the field of nuclear spins can take from hours to milliseconds or less. Many body interactions of electrons and
photons present us even with a wider range of time constants being limited in solids (such as a spin polarizer) only by the
plasma frequency. For a typical semiconductor, that frequency is around $10^{-14} s^{-1}$ corresponding to time constants down
to $10^{-14} s$. Naturally, if we include the frequencies derived from the mass of an electron and its energy, one has to
include possibilities of $10^{-20} s$ time constants and below. To all of these possibilities of time dependencies of the
functions $A, B$, one needs to add the fact that actual Aspect type of experiments are based on coincidence counting i.e. the
time correlation of the measurements in the two experimental wings play a major role. This leads to a true ``entanglement" of
the time dependent joint probabilities of measurement outcomes $A, B$. One might view this fact in a Shakespearean fashion: all
these random choices of settings are being made by the players, and time weaves then a pattern into it.

How, specifically, these reasonable space-time dependencies actually lead to the quantum result of Eq.~(\ref{11may25n2}) is
still subject of discussions.
EPR experiments that employ time-coincidences to identify pairs effectively apply a filter
to the data and it has been shown that such a filtering mechanism may lead to violations
of a Bell inequality, opening the route to a description in terms of locally causal, classical models~\cite{fine}.
A first concrete model of this kind was proposed by S. Pascazio who showed
that his model approximately reproduces the correlation of the singlet state~\cite{pasc}, violating Bell inequalities
Models that rigorously reproduce the results of quantum theory for the singlet
and uncorrelated state are given in Refs.~\cite{raed0,raed1,zhao,mich}.

Will these facts end the non-locality discussions? Probably not, because one can also remove the cyclicity with labels from
outside the light cone. The simplest choice is, of course, using $\lambda$'s that depend on the setting parameters $\bf a, b,
c...$ of the other sides. For Aspect type of experiments, that amounts to the use of space-time coordinates from outside the
light cone.  Our main point is, however, that the cyclicities can also be removed by space-time dependencies {\it within} the
light cone. The question is, of course, whether such time dependencies are reasonable and are indeed elements of a complete
physical theory. The answer to this question and the nature of the infinite set of space-time labeled functions as well as that
of the $\lambda$'s are, if they exist at all, currently not known. However, even with the lack of this knowledge, we can answer
an important question: can quantum mechanics be considered a complete physical theory?

\section{Kolmogorov vs quantum probability, completeness}

We first summarize the above findings. One can find different elements of a Boolean or $\sigma$-algebra for all possible events.
The elementary events can be understood by both, Einstein's definition of space-time events, and Kolmogorov's definition of the
events of a $\sigma$-algebra.  These events, corresponding to different experiments and their abstractions, do not lead to any
but the trivial Boole type inequalities such as Eq.~(\ref{11may23n3}) that can never be violated. Thus, at least in principle,
one can always find functions on a $\sigma$-algebra that describe any possible experimental result of macroscopic events that
can be recorded by our anthropomorphic methods. Because all these recordings can in that way be described without contradiction,
it follows that the averages of macroscopic indicator readings, that are described by quantum theory, can also be described
without contradiction. This answers the important question whether Kolmogorov probability can be made consistent with the
results of quantum mechanics. It always can be, because one always can find functions on a probability space that follow only
trivial Boole-Bell equations. If, however, the experimental results are not all denoted by different space-time coordinates, if
some are assumed to have the same coordinates or not to depend on coordinates, then nontrivial inequalities and contradictions
may arise. These contradictions call then for additional space-time labels or, speaking from the viewpoint of causal physical
theories, additional space-time dependencies.

To find a physically valid theory within the framework of Boole and Kolmogorov, it is therefore necessary to explore all
possible topological combinatorial cyclicities and to determine the necessary space-time labels, in order to arrive at a
contradiction free set of random variables that are defined as functions on a $\sigma$- algebra. Then one needs to select the
physically reasonable space-time connections that may underlie this $\sigma$-algebra. The advantage of quantum mechanics, as
compared to this cumbersome procedure, is that quantum mechanics works with a non-commutative algebra and, compensating for this
luxury, never attempts to explore the logical and causal space-time connections between single events or small sets of events.
Quantum theory only insists on agreement with averages of large numbers of experiments. If we wish to compare Boole's or
Kolmogorov's probability theory with quantum mechanics, and render judgements on their completeness with respect to a physical-science
point of view, the following fact should be considered. The introduction of probabilities is the hallmark of all of these
theories and does, by itself, not necessarily represent an incompleteness. It is indeed generally assumed that quantum mechanics
gives all the information about averages that we possibly can have. To achieve this result, however, quantum mechanics does
sacrifice the prediction of single and small sets of, events.

The elements of Kolmogorov's $\sigma$-algebra and functions of them must fulfill certain requirements not contained in
conventional quantum theory:  they must not contradict any facts following from the topological combinatorial cyclicities of
Vorob'ev. These rules for the data of experiments, therefore, impose physically speaking a ``law" on the space-time dependencies
of the outcomes. Because this law is not contained in quantum mechanics, but does relate to the physical measurement outcomes
that we macroscopically record, quantum mechanics exhibits an incompleteness in the sense of Einstein, Podolsky and Rosen.


\begin{thebibliography}{99}

\bibitem{bell} Bell J. S. in {\it The foundations of quantum mechanics}, Bell M., Gottfried K., and Veltman M. eds. (2001), World Scientific,
Singapore-New Jersey-London-Hong Kong.

\bibitem{boole} Boole G., Philos. Trans. R. Soc. London, 152 (1862) 225.

\bibitem{hmd} Hess K., Michielsen K. and De Raedt H., Europhys. Lett. {\bf 87} (2009) 60007.
\bibitem{hmd1} Hess K., Michielsen K. and De Raedt H., Europhys. Lett. {\bf 91} (2010) 40002.
\bibitem{hmd2} De Raedt H., Hess K., Michielsen K., J. Comp. Theor. Nanosci. {\bf 8} (2011) 1011.

\bibitem{cuisine} Bell J. S. in {\it Between Science and Technology} Sariemijn A. and Kroes, P. eds. (1990),
Elsevier Science publishers B. V. (North Holland), 97 -- 115.

\bibitem{EPR} Einstein A., Podolsky A. and Rosen N., Phys. Rev. {\bf 47} (1935) 777.

\bibitem{vaxjo} {\it Foundations of probability and physics - 3} Andrei Khrennikov, editor (2004), American Institute of Physics, AIP conference proceedings 750.

\bibitem{KHRE08} Khrennikov A., Theor. Math. Phys. {\bf 157} (2008) 1448.

\bibitem{KHRE09} Khrennikov A., {\it Contextual Approach to Quantum Formalism}, (2009), Springer, Berlin.

\bibitem{NIEU11} Nieuwenhuizen T., Found. Phys., {\bf 41} (2011) 580.


\bibitem{kolmo} Breuer H. P. and Petruccione F., {\it The theory of open quantum systems} (2002), Oxford University Press, pp. 3 -- 57.

\bibitem{vorob} Vorob'ev N.N., {\it Theory of probability and its applications} (1962) 147 -- 163.

\bibitem{bell1} see \cite{bell} pp. 7 -- 12

\bibitem{Asp} Aspect A., Dalibard J. and Roger G., Phys. Rev. Lett. {\bf 49} (1982) 1804.

\bibitem{lgarg} Leggett A. J. and Garg A., Phys. Rev. Lett. {\bf 54} (1985) 857.

\bibitem{mer} Mermin N. D., Europhys. Lett. {\bf 67} (2004) 693.

\bibitem{bvn} see \cite{bell} pp. 5

\bibitem{willi} Williams, D. (2001) {\it Weighing the Odds}, Cambridge University Press , Cambridge.

\bibitem{hp1} Hess K. and Philipp W., PNAS {\bf 98} (2001) 14228.

\bibitem{hp2} Hess K. and Philipp W., PNAS {\bf 101} (2004) 1799.

\bibitem{fine} A. Fine, Synthese 50, (1982) 279.
\bibitem{pasc} S. Pascazio, Phys. Lett. A 118, (1986) 47.
\bibitem{raed0} K. De Raedt, K. Keimpema, H. De Raedt, K. Michielsen, and S. Miyashita, Euro. Phys. J. B 53, (2006) 139.
\bibitem{raed1} H. De Raedt, K. De Raedt, K. Michielsen, K. Keimpema, and S. Miyashita, J. Comp. Theor. Nanosci. 4, (2007) 957.
\bibitem{zhao} S. Zhao, H. De Raedt, and K. Michielsen, Found. of Phys. 38, (2008) 322.
\bibitem{mich} K. Michielsen, F. Jin, and H. De Raedt, J. Comp. Theor. Nanosci. 8, (2011) 1052.


\end{thebibliography}
\end{document}